\documentclass[twocolumn,showpacs,preprintnumbers,a4paper,pre,notitlepage]{revtex4}
\usepackage{graphicx,float}
\usepackage[caption=false]{subfig}
\usepackage{amssymb, amsmath,amssymb,amsfonts}
\usepackage{dcolumn,lipsum}
\usepackage{bm}
\usepackage{color}
\usepackage[utf8]{inputenc}
\usepackage[english]{babel}
\usepackage{dsfont}

\begin{document}

\title{A minimally invasive neurostimulation method for controlling epilepsy seizures}

\author{Malbor Asllani, Timoteo Carletti}
\affiliation{naXys, Namur Institute for Complex Systems, University of Namur, rempart de la Vierge 8, B 5000 Namur, Belgium}

\begin{abstract} 
Many coordination phenomena in Nature are grounded on a synchronisation regime. In the case of brain dynamics, such self-organised process allows the neurons of particular brain regions to behave as a whole and thus directly controlling the neural activity, the muscles and finally the whole human body. However, not always such synchronised collective behaviour is the desired one, this is the case of neurodegenerative diseases such as Parkinson's or epilepsy where abnormal synchronisation induces undesired effects such as tremors and epileptic seizures. In this paper we propose an innovative, minimally invasive, control method able to effectively desynchronise the interested brain zones and thus to reduce the onset of undesired behaviour. 
\end{abstract}


\maketitle
\section{Introduction}

Synchronisation is one of the most important mechanism responsible for self-organisation in living beings~\cite{Pikovsky2003,Manrubia2005,Boccaletti2008}. Regular and periodic activity emerging from a collective behaviour of a set of interacting cells~\cite{strogatz,winfree} have been observed to be crucial for the operation of most processes of living organisms. Representative and relevant examples are the myocyte cells whose collective pulsing in unison makes possible the pumping of the blood from the heart~\cite{peskin}, or the neurons whose signals, sent in synchrony with each other, command our muscles with a impressive and somehow mysterious accuracy~\cite{spinal}. The neurons dynamics and more generally the brain one have attracted the interest of scientists in recent years, who increased their efforts to understand the underlying dynamics responsible for such emergent behaviour. It has been, for example observed that a lack of dopamine in the basal ganglia region of the brain, causes the Parkinson's disease symptoms such as uncontrolled and continuous tremors, rigidity and abnormal gait~\cite{Parkinson}. Even worse is the case of epilepsy where strong and often violent episodes of shaking, unexpectedly occur~\cite{epilepsy}. Independently from where the responsible neurons are located, in the depth part of the brain as in the case of Parkinson's disease, or in the cortex as in epilepsy, the common factor is that these cells abnormally synchronise, for reasons not yet completely clear~\cite{glass}. To tackle such issue is thus crucial to reduce/eliminate the synchronisation process to avoid these undesired effects and {control} the symptoms~\cite{notsync}. 

For the time being, the first used approach to treat the disease, is the administration of oral drugs whose result is partially effective over time for Parkinson's disease~\cite{Parkinson} but inefficient at all for nearly one-third of epileptic patients~\cite{epilepsy}. Starting from this setting, scientists have explored new clinical methods such as neurostimulation, which basically consists in an induced modulation of the neuronal activity in order to desynchronise the phase dynamics of neurons~\cite{tass,neurostim, neurocontrol,pyragas}. It can be invasive such as the Deep Brain Stimulation (DBS) where microelectrodes are inserted in the basal ganglia~\cite{DBS} or less invasive such as the Transcranial Magnetic Stimulation (TMS) where an external magnetic field interferes with the neuronal activity~\cite{neurostim}. 

Neurostimulation can be applied in other cases of severe mental disorders such as major depression or obsessive-compulsive disorder~\cite{neurostim} and it results to be the cutting edge in the fighting of neurological diseases especially when no other alternative treatment is effective.

In this paper, we will focus on the study of a novel minimally invasive
neurostimulation procedure principally oriented to suppress the epileptic
seizures although we believe it can also be applied to deal with other
diseases, as the ones  mentioned above. Based on the fact that neurons share
very similar dynamical properties among them and that they constitute a
strongly connected network of cells, one can infer that synchronisation
spontaneously and easily occurs; in other words, neurons are able to
synchronise even working in a weakly coupled regime, indeed the parameter
responsible for the interaction among neurons can, without any loss of
generality, be considered small. An evidence of this claim can be obtained using the paradigmatic Kuramoto model~\cite{kuramoto1975,kuramoto} to describe the behaviour of the network of linked neurons once one focuses on their phases, namely to consider each neuron as an oscillator, nonlinearly coupled with the other ones.

The proposed method is based on a theoretical control process we recently
presented~\cite{gjata} able to reduce the synchronisation state of
nonlinearly coupled neurons, modelled using the Kuramoto
model. In particular in the present work, we simplify such theoretical control term to make it operational, to be implementable in real cases and hence to be a valid alternative to {existing neurostimulation methods} because it is not as invasive as the latter. Roughly speaking, once the coupling 
among neurons is below a certain critical value, the system does not
self-organise neither seizures occur, accordingly the control, even if in action, is much smaller
compared to the interaction among the neurons and so it almost does not affect their
activity. On the contrary, once the coupling among the neurons is strong
enough to enhance the synchronisation phenomenon then the control gets stronger and induces a desynchronisation of the neurons dynamics with the consequent
suppression of the seizures. Keeping the {stimulus}, namely the
control parameter, in the brain cortex at its lowest possible value in
both the phase-unlocked and phase-locked regime is important for avoiding any
collateral effect such as hallucinations or hypersexuality, usually observed in
other neurostimulation methods due to exceeding {stimuli}~\cite{hyper}. For this reason, the proposed procedure for controlling the onset of the symptoms of
epilepsy is optimised to get the right compromise between reducing the seizures but at the
same time being as less invasive as possible.  

Let us observe that our method allows us also to provide a theoretical framework where other empirically determined control strategies proposed in the literature \cite{tass,neurocontrol,pyragas} can be grounded on solid theoretical bases.

The proposed method~\cite{gjata} is inspired by the Hamiltonian control
theory~\cite{vittot,chandre} whose validity has been already successfully
demonstrated in other domains~\cite{ciraolo,carletti}, and leans on the
Hamiltonian formulation of the synchronisation process recently proposed in~\cite{timme}. However,
this theoretical control procedure assumes a thorough knowledge of the system observables in terms of network topology and phase variable and more importantly all the interacting neurons should be directly
controlled. For sure, this setting doesn't apply for the brain where in the
best case, we can just measure the local dynamics and can interfere, using
{micro}electrodes, only  with a very limited number of (zones of) neurons compared
with the whole number of neurons involved. To tackle this problem, we hereby adapt
such theoretical control in order to limit the number of necessary {micro}electrodes
to achieve the desired level of control and at the same time restricting the
required information on the signal we measure from the same electrodes.  

In the following section we will introduce the mathematical formalism which
describes the synchronisation phenomenon. Then, we will give a short
presentation of the Hamiltonian control theory and we invite the interested reader to consult~\cite{gjata} 
to have more details. In section IV, we illustrate the application of the newly presented method
 to the neurons desynchronisation in the framework of the Kuramoto model, while in section V we extend the method to the more general Stuart-Landau model. We will then conclude by summing up our results.

\section{Neurons modelled as nonlinear oscillators}
\label{sec:themodel}

As already mentioned, the abnormal synchronisation of the neural activity observed in brain regions (thalamus, hippocampus, basal ganglia) is responsible for the symptoms (tremors, dystonia) emerging in neurological diseases such as Parkinson's disease, epilepsy, chronic pain, obsessive-compulsive disorder, just to mention few of them.
Despite the very different nature of the systems exhibiting synchronisation phenomena, most of the main features are quite universal and can thus be described using the paradigmatic Kuramoto model (KM)~\cite{kuramoto1975,kuramoto,syncopen,acebron2005,Arenas2008} of nonlinearly coupled oscillators, that in turns can be obtained from the more general Stuart-Landau \cite{ermentrout,pyragas} equation well suited to describe the normal form of a supercritical Andropov-Hopf bifurcation, namely a system able to switch back and forth from a stationary state into a periodic one - limit cycle - accordingly to a single bifurcation parameter:
\begin{eqnarray}
\dot{z}_k&=&(1+\j \omega_k-|z_k|^2)z_k+Z_k,\label{eq:stuart}\\
\mathrm{where}\;\; &Z_k&=\frac{K}{N}\sum_{j=1}^N A_{kj}z_j\nonumber
\end{eqnarray}
here the complex variable $z_k=\rho_k e^{\j\phi_k}$ encodes the information about the amplitude $\rho_k$ and the phase ${\phi_k}$ of the coupled oscillators, $\j=\sqrt{-1}$ is the imaginary unity, $\omega_k$ are the natural frequencies of the oscillators and are drawn from a symmetric unimodal distribution $g(\omega)$, $K$ is the coupling strength and the symmetric adjacency matrix $A_{kj}$ encodes the connections among the $N$ oscillators, $A_{kj}=A_{jk}=1$ if oscillators $k$ and $j$ are directly coupled and zero otherwise. Considering the real part of Eq.~\eqref{eq:stuart} and assuming the amplitudes to be almost equal, $\rho_k\sim \rho_j$ for all $k$ and $j$~\footnote{{This statement is true in a weakly coupled regime as the case of a neuron ensemble.}}, we eventually obtain the Kuramoto model 
\begin{equation}
\dot{\phi}_k = \omega_k + \frac{K}{N}\sum_{j=1}^N A_{kj}\sin(\phi_j-\phi_k)\, .
\label{eq:kurmod}
\end{equation}
Let us remind that the classic Kuramoto model corresponds to an all-to-all coupling~\cite{kuramoto1975}, $A_{kj}=A_{jk}=1$, for all $k\neq j$, $A_{kk}=0$. The model can be rewritten using the order parameter~\cite{kuramoto}
\begin{equation}
Re^{\j\Psi} = \frac{1}{N} \sum_{j=1}^N e^{\j\phi_j}\, ,
\label{eq:par_ord}
\end{equation}
a macroscopic index to measure the strength of the synchronisation, if $R\sim 0$, the oscillators are almost independent each other while if $R\sim 1$ they are close to phase-lock. Substituting the above definition in the original model we get the mean-field equation
\begin{equation}
\dot{\phi}_k = \omega_k + KR\sin(\Psi-\phi_k)\, .
\label{eq:kur_MF}
\end{equation}
Thus the oscillators are no longer directly coupled to each other, but to the global oscillator with phase $\Psi$. 

\section{Hamiltonian control and the synchronisation problem}

The KM is a dissipative system, however an $N$ dimensional Hamiltonian system $H(\boldsymbol{\phi},\mathbf{I})$ written in angles variables $\boldsymbol{\phi}=(\phi_1, \dots \phi_N)$  and actions variables $\textbf{I}=(I_1, \dots, I_N)$, have been recently proposed~\cite{timme} that embeds as particular orbits the ones of the KM, more precisely one can define the invariant Kuramoto torus $\mathcal{T}:=\{ (\mathbf{I},\boldsymbol{\phi})\in\mathbb{R}^N_+\times \mathbb{T}^N: I_i=1/2 \; \forall i\}$ and prove~\footnote{We refer the interested reader to~\cite{timme,gjata} 
, to have a more detailed description of the model and of its properties.} that the restriction of time evolution of the angles variables $(\phi_1, \dots \phi_N)$ to this torus coincides with Eq.~\eqref{eq:kurmod}.

In~\cite{timme}, authors have numerically shown and analytically proved that when the Kuramoto oscillators enter in a synchronisation state, then the dynamics of the actions close to the Kuramoto torus, become unstable and exhibit a chaotic behaviour. Based on this result, our aim is to reduce the synchronisation in the KM~\eqref{eq:kurmod} by controlling the Hamiltonian system $H(\boldsymbol{\phi},\mathbf{I})$ by adding a small control term able to increase the stability of the invariant torus $\mathcal{T}$, hence to reduce the chaotic behaviour close to such torus, and thus to impede the phase-lock of the coupled oscillators. Let us rewrite the Hamiltonian in the form $H=H_0+V$, where $H_0$ is the integrable part, i.e. the uncoupled harmonic oscillators, and $V$ the non-linear term, namely the $KR\sin(\Psi-\phi_k)$ function in the KM, that can be considered as a perturbation of $H_0$ because of the small parameter $K$. Then, roughly speaking, the main idea of Vittot and coworkers~\cite{vittot,ciraolo} is to add to $H$ a small control term $f\sim \mathcal{O}(K^2)$, whose explicit form depends on $V$, in order to reduce the impact of the perturbation $V$, as previously said to increase the stability of the invariant torus. The size of $f$ implies that the controlling procedure is much less invasive than other techniques generally used in control theory and also able to give a prompt response to possible abnormal dynamics without time lags and more importantly, without any need for further measure of the system state. Assuming a technical condition on the natural frequencies~\footnote{Let us observe that the theory by Vittot can also handle more general cases where such additional assumption is relaxed.}, namely $\boldsymbol{\omega}=(\omega_1,\dots,\omega_N)$ to be not resonant, i.e. for all $\textbf{k}\in\mathbb{Z}\setminus\{0\}$ then $\textbf{k}\cdot \boldsymbol{\omega} \neq 0$, one can straightforwardly compute the required control term $f(\boldsymbol{\phi},\mathbf{I})$.

The embedding of the KM into the Hamiltonian system is based on the existence of the invariant torus $\mathcal{T}$, which is no longer invariant for the controlled Hamiltonian $H_0+V+f$, {nevertheless it is possible to} provide an effective control by truncating the latter to its first term, such that the resulting controlled Hamiltonian system preserves the Kuramoto torus. One can thus transpose this information into the KM and achieve a control strategy:
\begin{equation}
\dot{\phi}_k = \omega_k + KR\sin(\Psi-\phi_k)+h_k(\phi_1,\dots,\phi_N)\, ,
\label{eq:kur_MF_control}
\end{equation}
where $h_k(\phi_1,\dots,\phi_N)$ is the contribution of the control $f$ to the angles dynamics and it is explicitly given by:
\begin{widetext}
\vspace*{-.55cm}
\begin{align}
h_k(\phi_1,\dots,\phi_N) = &-\frac{K^2}{4 N^2}\left[\sum_j A_{kj}\cos(\phi_j-\phi_k)\sum_l\frac{A_{kl}}{\omega_l-\omega_k}\cos(\phi_l-\phi_k){+}\sum_j\frac{A_{kj}}{\omega_j-\omega_k}\sin(\phi_j-\phi_k)\sum_l A_{kl}{\sin}(\phi_l-\phi_k)+\right.\nonumber\\&\left.-\sum_l\left(A_{kl}\cos(\phi_k-\phi_l)\sum_j\frac{A_{jl}}{\omega_j-\omega_l}\cos(\phi_j-\phi_l){+}\frac{A_{kl}}{\omega_k-\omega_l}\sin(\phi_k-\phi_l)\sum_j A_{jl}{\sin}(\phi_j-\phi_l)\right)\right]\, ,
\label{eq:kurcontrol}
\end{align}
\vspace*{-.2cm}
\end{widetext}
where with a slight abuse we used the same letter to denote the new angular variable, namely the controlled one. The truncation to the first order of the control term $f$, finds its justification in the Hamiltonian perturbation theory, moreover the approximation is better the smaller the perturbation parameter $K$, being $f\sim\mathcal{O}(K^2)$.

To simplify the previous equation, let introduce a second modified local order parameter, that depends now on the node index:
\begin{equation}
\tilde{R}_ke^{\j\Psi_k} = \frac{1}{N} \sum_{j=1}^N \frac{e^{\j\phi_j}}{\omega_j-\omega_k}\, .
\label{eq:par_ord_2}
\end{equation}
Then a straightforward computation allows to rewrite the control term, under the hypothesis of all-to-all coupling as:
\begin{equation}
h_k(\phi_1,\dots,\phi_N) = -\frac{K^2}{4}\left[ R\tilde{R}_k\cos(\Psi-\Psi_k)-\mathcal{B}_k\right]\, ,
\label{eq:phictrlcompl}
\end{equation}
where the term $\mathcal{B}_k$ is defined by
\begin{eqnarray*}
\mathcal{B}_k&=&\frac{1}{N}\sum_{l}\cos(\phi_k-\phi_l)\cos(\Psi_l-\phi_l)\tilde{R}_l\notag\\&+& \sum_l \frac{\sin(\phi_k-\phi_l)}{\omega_k-\omega_l}\sin(\Psi-\phi_l)R\, ,
\end{eqnarray*}

\section{Effective desynchronisation of the phases of coupled neurons}

Before we enter into the technical details of the proposed method, let us first comment on the obtained analytic result and discuss about its advantage with particular attention to the control of the epileptic seizures. As already anticipated earlier in this paper, our principal aim is to develop a novel method which targets the desynchronisation of the neurons of the brain cortex responsible for causing the seizures. However, since the neurostimulation technique suitable for such purpose is often strongly invasive, our aim is to optimise the control strategy by letting the latter to act only when necessary, more precisely the control should dynamically \lq\lq switch on\rq\rq when the seizures start to appear and again dynamically \lq\lq turn off\rq\rq during the normal neuronal regime, namely it should be very small in intensity. This is actually what the proposed control term~\eqref{eq:kurcontrol} does; the two main contributions to the control are the prefactor $K^2$ and the denominators containing the differences of natural frequencies $\omega_j-\omega_k$, which is of the order of the width of the frequencies distribution, $g(\omega)$. Because it is well known that the critical value of the coupling strength $K_c$ ($<1$) is of the order $\sim g(\omega)$, the control term becomes of order $K$ in the critical regime, namely once it is necessary to reduce the synchronisation, on the other hand during the normal regime, the control size is much smaller than the critical one, $K^2\ll K_c$, and in consequence the method is minimally invasive.
\begin{figure*}
\center
\includegraphics[width=.85\textwidth]{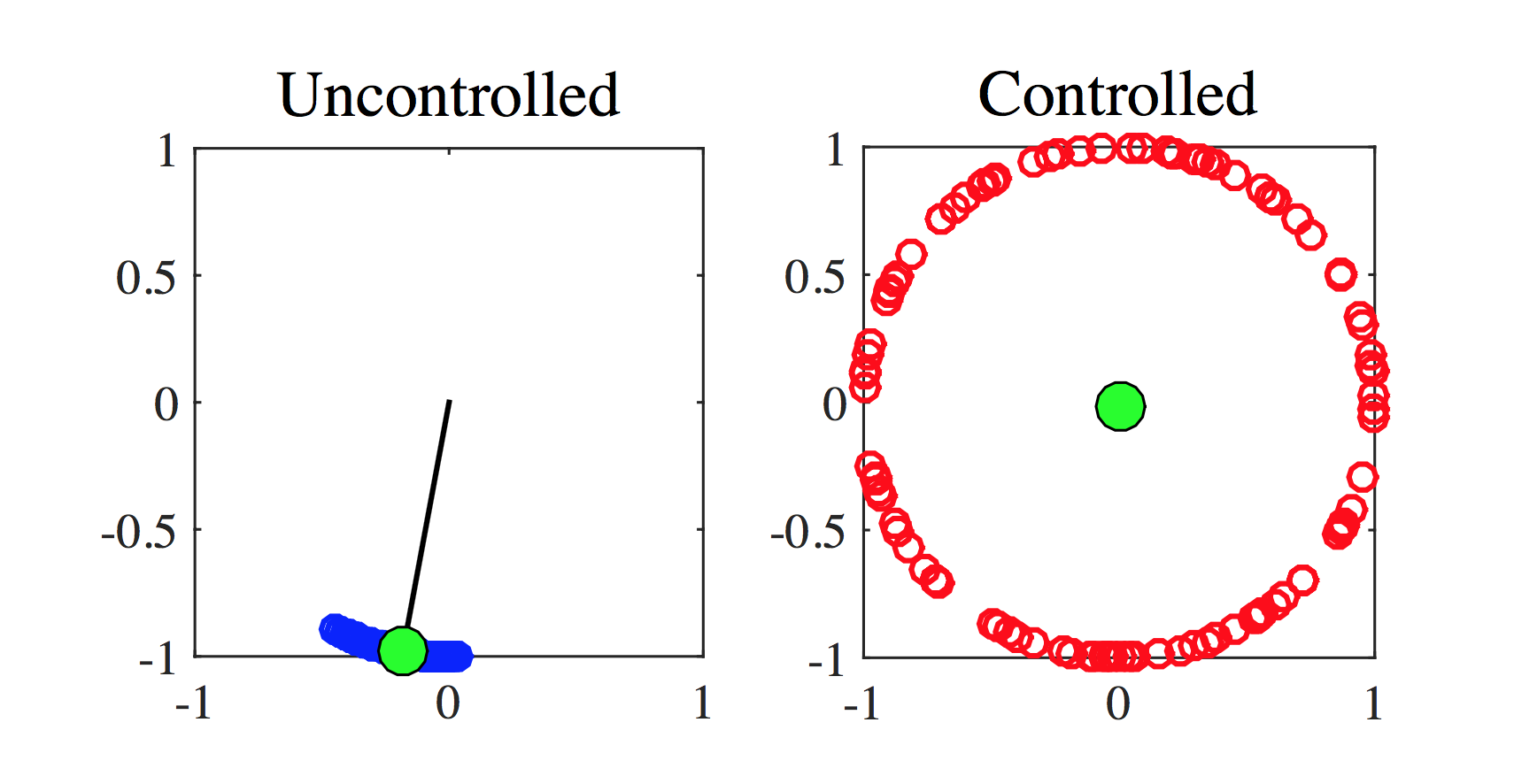}
\vspace*{-.5cm}
\center{\hspace*{.75cm}{\Large $a)$}\hspace*{6.2cm} {\Large $b)$}}
\caption{A snapshot of the Kuramoto dynamics at a generic time. $N=100$ oscillators (circles) are drawn on the unitary circle, their angular position is given by the oscillator phase. The dynamical behaviour presented in panel a) corresponds to the uncontrolled phase-locked regime for a coupling parameter $K=0.5$. In panel b) we report, for the same coupling parameter, the controlled case obtained acting on $M=30$ oscillators, resulting in a desynchronised behaviour. The underlying network is a Newman-Strogatz small-world network~\cite{newman} with parameter $p=0.85$. The green circle represents the Kuramoto order parameter, its angular position is given by the angle $\Psi$ while its distance from the origin is $R$.}
\label{fig:Fig1}
\end{figure*}

Let us now come back to the theoretical control term~\eqref{eq:phictrlcompl} and prove that one can realise it as an operational control strategy. The first observation is that the latter requires the control of all unities, neurons or neural zones, and this is impossible to be done in a realistic situation. The second observation is that the control demands the exact knowledge of the topology all the interacting involved cells. From the practical point of view we \textit{a priori} know that in order to control the synchronisation, we should interfere with the neuronal dynamics, by sending an electrical signal through a {micro}electrode inserted into a suitable zone of the brain tissue. For this reason the main dilemma inherent with neurostimulation methods is how to be as less invasive as possible but at the same time as efficient as possible? To give a possible solution to this issue we will try to mathematically carve the formula~\eqref{eq:phictrlcompl} to fit our goal of having an operational control. We hereby anticipate that we will able to achieve our goal, by obtaining a desynchronisation effect using a limited number of controllers ({micro}electrodes) as good as the one involving the control of all the neurons. Regarding to the network of interacting neurons we will work under the hypothesis that the latter is a strongly connected one; this assumption is justified by experimental observations~\cite{sporns} which classify the brain networks as \textit{small-world} ones, a topology which particularly promotes the strong connection between the individuals belonging to such networks \cite{watts}. Moreover we can safely use a all-to-all coupling without substantially modifying the resulting dynamics. 

The first empirical observation is that the second term, $\mathcal{B}_k$, in Eq.~\eqref{eq:phictrlcompl} is very often much smaller than the first one, mathematically this fact can be understood because this term involves averages of products of oscillatory functions that can thus compensate each other. The second observation is that one can hardly compute the local phase $\Psi_k$ using a limited number of {micro}electrodes sampling few neurons, we then decide to replace the latter with the neuron phase $\phi_k$. We are aware of impact of these working assumptions, nevertheless the justification of these choices is obtained {\it a posteriori} by observing that the effective control performs very well. In conclusion the proposed local control strategy is given by:
\begin{equation}
h_k(\phi_1,\dots,\phi_N) = -\frac{1}{4}\gamma K^2R\tilde{R}_k\cos(\Psi-\phi_k)\, ,
\label{eq:kurcontrol_2}
\end{equation}
let us observe that we added a free parameter $\gamma$ to take into account the not perfectly known network structure, in particular it can be set equal to the ratio of the average connectivity with the possible maximum number of links, which is a macroscopic parameter probably known with good precision in advance. In conclusion let us observe that the local control term is built using a cosine function which is nothing but the coupling term in the KM~\eqref{eq:kurmod} delayed by a quarter of its period $T$, we thus here recover the empirical rule proposed by \cite{tass}. The operational control of a given neuron goes thus as follows: compute the signal from a given neuron through a {micro}electrode, delay the signal by $T/4$, multiply it by $\gamma K\tilde{R}_k$, where $\tilde{R}_k$ is computed using a limited number of signals from neurons where {micro}electrodes are inserted, and re-inject the new signal in the initial neuron. In this way the latter will desynchronise and break away from the group of neurons acting as a single giant oscillator. 

We however observe that this process is not enough to desynchronise the whole system, only the controlled neuron where the {micro}electrode is placed is desynchronised and because we want to limit the number of implanted electrodes we cannot sufficiently reduce the {symptoms} using this strategy. To achieve our goal, it is necessary to indirectly influence the behaviour of the other neurons and this can be done considering the fact that the same {micro}electrode that control a given neuron, is able to produce a non invasive electromagnetic field potential~\cite{Nunez, Rinzel}. To be more specific, let us denote with $S_k^{stim}$ the stimulation signal generated on the $k$-th neuron by the potential produced by the electrode located at cell $l$-th, namely the control term:
\begin{equation}
S_k^{stim}=c_s\sum_{l=1}^M e^{-2r_{kl}}h_l(\phi_1,\dots,\phi_N)\, ,
\label{eq:stim}
\end{equation}      
where $r_{kl}$ and $c_s$ are respectively, the distance of the interested node $k$ from the origin of the electromagnetic field $l$, and the strength of the potential which in our case is taken $c_s=1$, finally $M\ll N$ is the number of controlled neurons, namely of implanted {micro}electrodes. 
\begin{figure*}
\includegraphics[width=1\textwidth]{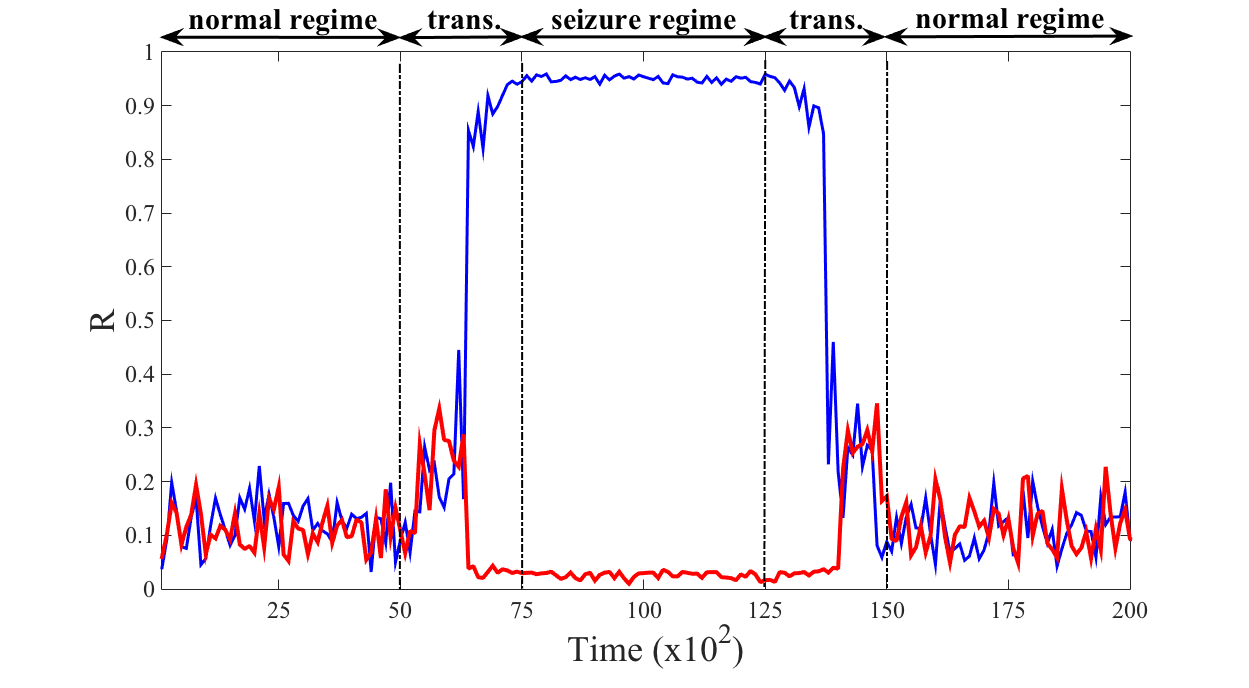}
\caption{Onset of an epileptic seizure in the Kuramoto-like neuron population and the outcome of the controlled system. We represent the order parameter $R$ (blue curve for the KM model and red curve for the controlled one) as a function of time for $100$ coupled oscillators linked using a Newman-Watts small world network~\cite{newman} and $30$ {micro}electrodes for the controlled case. To smooth the results, each curve is the average over $20$ independent realisations. We assume $K$ to be small in the interval $[0,5000]$ fluctuating around the average value $0.1$, during this period of time both systems behave similarly and do not exhibit synchronisation. Then we assume the coupling parameter to start to increase, $[5000,7500]$, to eventually remain, $[7500,125000]$, quite large, $0.5$ on average; we can observe that the KM falls in a synchronised state while the controlled one still exhibit a non-synchronised one. Once the coupling parameter decreases and fluctuates again around a small value, $0.1$, both systems recover the same non-synchronised state.}
\label{fig:Fig2}
\end{figure*}

In Fig.~\ref{fig:Fig1} we report the results for a generic numerical simulations of the Kuramoto model, oscillators are represented as circles laying on the unitary circle whose angular coordinate is given by the oscillator phase, the green circle identifies the Kuramoto order parameter, its angular position is given by $\Psi$ while the distance from the origin, the black segment (clearly visible on the panel a)), represents $R$. Let us observe that the longer such segment, namely the larger $R$, the stronger is the synchronisation of the oscillators, as it can clearly be appreciated on the panel a) where most of the circles are very close to the green one. On the other hand (see panel b)) one can observe that in the case of non-synchronisation the oscillators are quite uniformly distributed on the circle, resulting thus in $R\sim 0$. The network chosen for coupling of the $100$ oscillators is a Newman-Watts small world one~\cite{newman}. We can clearly observe that in original, namely uncontrolled, Kuramoto model, for the chosen coupling parameter $K=0.5$ larger than the critical one $K_c\approx 0.4$, the oscillators tend to synchronise, they almost have the same phase (see panel a)); on the other hand, for the same value of the coupling parameter, but applying the effective control using $M=30$ oscillators, the behaviour is completely different, the oscillators almost uniformly distribute on the unitary circle (see panel b)) corresponding to a  desynchronised system.

To mimic the onset of an epileptic seizure in the brain and the action of the proposed control we realise the following numerical experiment using $N=100$ neurons connected using a Newman-Watts small-network, firstly without control and then controlled using $M=30$ {micro}electrodes. In both cases, during a given period of time, $[0,5000]$, we numerically solve the KM with a small control parameter fluctuating in time to mimic the physiological fluctuations one can observe in the brain; more precisely every $\Delta t=100$ time units, we draw a value for $K$ from an {uniform} distribution with support $[0.05,0.15]$, and thus with average $0.1$, and we follow the model dynamics with such fixed value for the control parameter during $\Delta t$ time units. Let us observe that the coupling parameter is smaller than the critical one, $K_c\approx 0.4$, and thus the system remains in a non-synchronised state as one can appreciate from Fig.~\ref{fig:Fig2} where we plot the order parameter $R$ as a function of time for the uncontrolled (blue) and controlled (red) KM. Observe moreover that both system behave very similarly (the curves are very close), hence the control, even if present, is not changing the dynamics when not needed.

Then we assume the coupling parameter to quickly increase and then to fluctuate around a large value, mimicking the fact the because of the disease the brain is not able to keep the normal behaviour; mathematically we assume that during the time interval $[5000,7500]$ every $\Delta t=100$ time units, we draw a value for $K$ from a {uniform} distribution whose average grows linearly in time from $0.1$ at $t=5000$ to reach $0.5$ for $t=7500$, while in the interval $[7500,125000]$ the coupling parameter is drawn from a {uniform} distribution with support $[0.55,0.65]$, and thus with average $0.6$. The results of the numerical simulations are striking, after a short transient time the uncontrolled system (blue curve) almost fully synchronises, $R$ is very close to $1$, while {the} controlled one remains in the non-synchronised phase, $R$ very close to $0$. Of course when the coupling parameter start to decrease to eventually reach again a small average value, the original system and the controlled one both exhibit a non-synchronised behaviour.

\section{Effective desynchronisation of the coupled neurons}

In the previous section we built an operational control able to effectively reduce the synchronisation onset in the phases of the neurons described by the Kuramoto model even for large values of the coupling parameter. Because the latter model can be derived (see Section~\ref{sec:themodel}) from the more general Stuart-Landau one, we naturally wonder if one could define a control strategy acting directly on the Stuart-Landau system self-consistently defined in terms of the control used for the Kuramoto model.

Let us consider Eq.~\eqref{eq:stuart} under the assumption of all-to-all connection with the additional term $Z_k^{ctrl}=-\j \frac{K}{4}\tilde{R}_kZ$, where $Z=\frac{K}{N}\sum_{j=1}^N z_j$, namely
\begin{equation}
\dot{z}_k=(1+\j \omega_k-|z_k|^2)z_k+Z+Z_k^{ctrl}\, ,
\label{eq:stuartctrl}
\end{equation}
A straightforward computation allows to show that the time evolution of the phase of the complex variable $z_k=\rho_k e^{\j \phi_k}$ in the previous system reduces to the controlled Kuramoto, see Eq.~\eqref{eq:kur_MF_control} where $h_k$ is given by Eq.~\eqref{eq:kurcontrol_2}, once we assume the amplitudes to be very close each other.

\begin{figure*}
\includegraphics[width=1\textwidth]{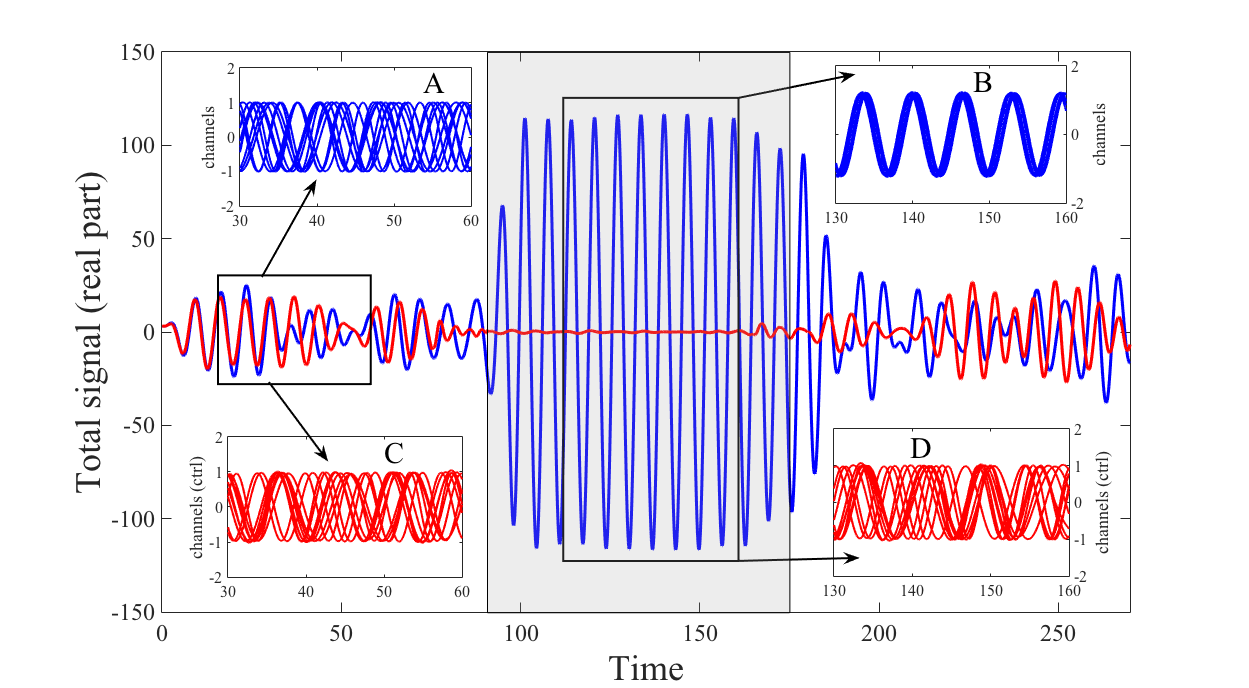}
\caption{Onset of an epileptic seizure in the Stuart-Landau-like neurons population and the outcome of the controlled system. In the main plot, we represent the (real part of the) total signal $\sum_k \rho_k \cos\phi_k$  (blue curve for the SL model and red curve for the controlled one) as a function of time for $100$ coupled oscillators coupled by a Newman-Watts small world network~\cite{newman} and $30$ {micro}electrodes for the controlled case. We assume $K$ to be small, namely fluctuating around the average value $0.1$, in the interval $[0,\sim 90]$; during this period of time both systems behave similarly and do not exhibit synchronisation (see inset A for the SL model and inset C for the controlled SL model). Then we assume the coupling parameter to start to increase to eventually remain quite large, on average $0.5$, in the time interval $[\sim 90,\sim 170]$; we can observe that the SL synchronises (see inset B) while the controlled one still exhibit a non-synchronised regime (see inset D). Once the coupling parameter decreases and fluctuates again around a small value, $0.1$, both systems recover the same non-synchronised state (data not shown).}
\label{fig:Fig3}
\end{figure*}

To prove the effectiveness of this control we perform a numerical simulation similar to the one proposed in the previous section to mimic the onset of an epileptic seizure but this time using the full Stuart-Landau model. More precisely we consider a system of $N=100$ neurons described by the Stuart-Landau model~\eqref{eq:stuart} and its controlled version~\eqref{eq:stuartctrl} ($M=30$ {micro}electrodes) using again a Newman-Strogatz network; initially the coupling parameter fluctuates around a small value and afterward it becomes larger. In Fig.~\ref{fig:Fig3} we represent the total (real part of the) signal, namely $\sum_k \Re z_k=\sum_k \rho_k \cos\phi_k$, for the original Stuart-Landau model (blue curve) and the controlled one (red curve). In the interval $[0,\sim 90]$, $K$ is small and both systems do not synchronise as it can be appreciated looking at the insets A (real part of the signal for $10$ generic neurons for the original Stuart-Landau model) and C (real part of the signal for $10$ generic neurons for the controlled Stuart-Landau model), the amplitude of the total signal is thus quite small. On the other hand for larger times, $[\sim 90,\sim 170]$ (roughly corresponding to the shaded central rectangular part of the figure), $K$ is larger and the Stuart-Landau system enters in a synchronised state (see inset B where again we plot the real part of the signal for the same $10$ generic neurons of inset A) while the controlled system remains in a non-synchronised state (see inset D where again we plot the real part of the signal for the same $10$ generic neurons of inset C), corresponding to a quite large amplitude for the total signal because now the amplitudes of each single signal do coherently add together.

\section{Conclusions}

In this paper, we have developed a new method to control the abnormal synchronisation of neuronal activity based on the Hamiltonian control formalism, applied to the paradigmatic Kuramoto model. We focused on the phase dynamics which lays the foundation for most of the basic functioning of the brain regions. As well-known when the coupling strength $K$ exceeds a critical value the phases of the electrical currents of the interested set of neurons get locked and as a result due to a resonance effect, the neural signal amplifies directly affecting the body activity. However, sometimes this behaviour is not the suitable one and can be associated to neurological diseases. This is the case for example of the epileptic seizures, strong uncontrolled rhythmic skeleton muscles which appears unexpectedly and may be dangerous. Often drugs are not sufficient to control (reduce) the seizures and external brain stimulation becomes necessary. Starting from this setting, we have proposed an efficient and minimally invasive control technique aimed to prevent the phase-locking and thus suitable for all the cases where the synchronisation is the responsible for undesired negative effects, as is the case of neurological diseases such as epilepsy, Parkinson's disease, severe mental disorders, etc. 

Starting from a theoretical results~\cite{gjata} we had further developed this control term and adapted it for realistic applications such as epilepsy. The main idea is to effectively control the interested brain region by reducing at most the collateral effects, namely to have few {micro}electrodes implanted which moreover are active only "when needed", namely the amplitude of the control term becomes comparable to the neuronal activity only when the latter enters in a abnormal synchronised state.
 
Starting from this control we have been able to define a control strategy acting directly on the Stuart-Landau model widely used to describe the interaction of coupled neurons~\cite{Deco2016} and numerically proved its effectiveness in suppressing the synchronisation state and thus the neuronal disease.


\begin{thebibliography}{1}
\bibitem{Pikovsky2003} A. Pivkosky, M. Rosenblum and J. Kurths, \textit{synchronization: A Universal Concept in Nonlinear Sciences}, Cambridge University Press (2003).
\bibitem{Manrubia2005} S.C. Manrubia, A.S. Mikhailov and D.H. Zanette, \textit{Emergence of Dynamical Order: Synchronization Phenomena in Complex Systems}, World Scientific (2005).
\bibitem{Boccaletti2008} S. Boccaletti, \textit{The Synchronized Dynamics of Complex Systems}, Elsevier (2008).
\bibitem{strogatz} S.H. Strogatz, \textit{Sync : The Emerging Science of Spontaneous Order}, Hyperion (2003).
\bibitem{winfree} A. T. Winfree, \textit{J. Theor. Biol.} \textbf{16}, 15 (1967).
\bibitem{peskin} C. S. Peskin, \textit{Mathematical aspects of heart physiology}, New York, CPAM (1975).
\bibitem{spinal} R. Berg, A. Alaburda and J. Hounsgaard, \textit{Science}, \textbf{315}, 390 (2007).
\bibitem{Parkinson} S. Sveinbjornsdottir, \textit{J. Neurochem.} \textbf{139}, 318 (2016).
\bibitem{epilepsy} B. Chang and D. Lowenstein, \textit{N. Engl. J. Med.} \textbf{349}, 1257 (2003).
\bibitem{glass} L. Glass, \textit{Nature} \textbf{410}, 277 (2001).
\bibitem{notsync} V. H. P. Louzada, N. A. M. Ara\'ujo, J. S. Andrade, Jr. and H. J. Herrmann, \textit{Sci. Rep.} \textbf{2}, 658 (2012).
\bibitem{neurostim} M. Hallet, \textit{Nature} \textbf{406}, 147 (2000).
\bibitem{tass} P. Tass, \textit{Prog. Theor. Phys. Suppl.} \textbf{150}, 281 (2003).
\bibitem{neurocontrol} O. Popovych and P. Tass, \textit{Front. Neurol.} \textbf{5}, 268 (2014).
\bibitem{pyragas} K. Pyragas, O. Popovych and P. A. Tass, \textit{EPL}, \textbf{80}, 40002 (2007). 
\bibitem{DBS} M. Kringelbach, N. Jenkinson, S. Owen and T. Aziz, \textit{Nat. Rev. Neurosci.} \textbf{8}, 623 (2007).
\bibitem{kuramoto1975} Y. Kuramoto, Lect. Notes in Physics \textbf{30}, 420 (1975).
\bibitem{kuramoto} Y. Kuramoto, \textit{Chemical Oscillations, Waves, and Turbulence}, New York, Springer-Verlag (1984).
\bibitem{gjata} O. Gjata, M. Asllani, L. Barletti and T. Carletti, \textit{Phys. Rev. E} \textbf{95}(2), 022209 (2017).
\bibitem{hyper} D. Burn and A. Tröster, \textit{J Geriatr Psychiatry Neurol.} \textbf{17}, 172 (2004).
\bibitem{vittot} M. Vittot, \textit{J. Phys. A: Math. Gen.} \textbf{37}, 6337 (2004).
\bibitem{chandre} C. Chandre and et. al., \textit{Phys. Rev. Lett.} \textbf{94}, 074101 (2005).
\bibitem{ciraolo} G. Ciraolo et. al., \textit{Phys. Rev. E} \textbf{69}, 056213 (2004).
\bibitem{carletti} J. Boreux et. al., \textit{Int. J. Bifurcation Chaos} \textbf{22}, 1250219 (2012).
\bibitem{timme} D. Witthaut and M. Timme \textit{Phys. Rev. E} \textbf{90}, 032917 (2014).
\bibitem{syncopen} S.H. Strogatz, \textit{Physica D} \textbf{143}, 1 (2000).
\bibitem{acebron2005} J.A. Acebr\' on et al, Rev. Mod. Phys. \textbf{77}, 137 (2005)
\bibitem{Arenas2008} A. Arenas et al, Phys. Rep. \textbf{469}, 93 (2008).
\bibitem{ermentrout} B. Ermentrout and N. Kopell, \textit{J. Math. Biol.}, \textbf{29}, 195 (1991).
\bibitem{sporns} R. Bullmore and O. Sporns \textit{Nat. Rev. Neurosci.} \textbf{10}, 186 (2009).
\bibitem{watts} D. Watts and S. Strogatz \textit{Nature} \textbf{393}, 440 (1998).
\bibitem{Nunez} P. Nunez, \textit{Electrical fields of the brain}, Oxford University Press (1981).
\bibitem{Rinzel} J. Rinzel and G. Ermentrout, \textit{Analysis of the neural excitability and oscillations}, in C. Koch and I. Segev (eds.), Methods in Neuronal modelling from synapses to networks, MIT Press, pp. 135--169 (1989).
\bibitem{newman} M. Newman and D. Watts, \textit{Phys. Rev. E}, \textbf{60}, 7332 (1999).
\bibitem{Deco2016} G. Deco, M.L. Kringelbach, V.K. Jirsa and P. Ritter, preprint, \textit{http://biorxiv.org/content/early/2016/07/22/065284}, (2016)
\end{thebibliography}
\end{document}